\documentstyle[12pt,aaspp4]{article}
\def\etal{{\sl~et\,al.~}}

\begin{document} 

\title
{Discovery of Red Selected Arcs at z=4.04 behind Abell 2390}

\author
{Brenda Frye and Tom Broadhurst}
\affil
{Department of Astronomy, University of California, Berkeley, CA  94720}


\begin{abstract}

  We describe the properties of a system of red arcs discovered at
z=4.04 around the cluster A2390 ($z=0.23$).  These arcs are images
of a single galaxy, where the lensing is compounded by an elliptical
cluster member lying close to the critical curve of the cluster. The
combined magnification is estimated to be $\sim 20$, using HST images,
depending the gradient of the model potentials, implying an unlensed
magnitude for the source of $I_{AB}=25.5$. Keck spectroscopy reveals a
continuum well fitted by B-stars and attenuated by the Lyman-series
forest with an opacity consistent with z$\sim$4 QSO spectra, making
the arcs relatively red. Damped Ly$\alpha$ absorption is
observed at the source redshift with a relatively high column,
n$_{HI}=3\times 10^{21}$.  Curiously, Ly$\alpha$ emission is
spatially separated ($\sim 0.5$kpc) from the bright stellar continuum
and lies $\sim$ 300Km/s redward of interstellar absorption lines at
z=4.04. Similar redward shifts are found in all high redshift
galaxies with good spectroscopy, indicating that outward flows of
enriched gas are typical of young galaxies.  Finally, we briefly
comment on the notorious ``straight arc'' in A2390, which is resolved
into two unrelated galaxies at z=0.913 and z=1.033.
 
\end{abstract}


\section
{Introduction}

   Magnification by galaxy clusters is responsible for the discovery
of many distant galaxies. A clear example is the mirror symmetric arc
straddling the critical curve of A2218 (Kneib \etal 1996) which is
found to lie at $z=2.51$ (Ebbels \etal 1996). A much brighter object
``CB58'' with $V=20$ and $z=2.72$, was discovered in a redshift survey
of the cluster MS1512+36 (Yee \etal 1996). This object does not form
an obvious arc but appears to be an example of a fold image of an
intrinsically small source (Seitz \etal 1996) and hence near maximally
magnified for its size, so that it lies obliquely across the critical
curve.  More distant galaxies were uncovered by Trager \etal 1997, at
redshifts $3.34$ and $3.98$, behind the core of a more distant cluster
A851, $z=0.41$, (Dressler and Gunn 1992), in a study of irregular
cluster members. A flat central potential is the most reasonable
explanation here, generating similar tangential and radial stretching,
so the effect of lensing is mainly an enlargement of images with
little change in shape, not inconsistent with the weak distortions
claimed in the field of this cluster (Seitz \etal 1996). For
intermediate redshift clusters like this, the magnification of distant
galaxies is maximal for a flat potential. Most analogous to the lensed
galaxy presented here are the red arcs discovered by Franx \etal (1997) 
at $z=4.92$, behind the cluster CL1358+62 ($z=0.33$). Here the
magnification of the main arc is compounded by the potentials of the
cluster and an elliptical member.

  In this letter we describe a lensed system of arcs at $z=4.04$,
discovered behind the massive lensing cluster A2390 ($z=.231$) 
in a redshift survey of faint red objects behind
massive clusters.  We demonstrate that these arcs are all images of
one highly magnified galaxy and that their spectra are attenuated by
the the expected level of foreground Lyman-forest absorption. We find
evidence for gas outflow in common with both other high redshift
galaxies and well studied local starburst galaxies. This object is the
first very distant galaxy discovered in our general survey of
red cluster background objects.

\section {Observations}
 
  HST images of A2390 were used to select galaxies redder than the
cluster sequence for spectroscopy with the Keck. The imaging
observations (courtesy of B. Fort) comprise four long exposures
totaling 160 minutes in each of the F555W and F814W bands,
corresponding roughly to V and I. These images were aligned to the
nearest integer and then cosmic rays rejected separately in each
passband. Figure 1 (see Plate) shows a color representation of the
center of cluster. A striking lensing pattern is visible, including
the famous straight arc (Mellier 1989, Pell\'o \etal 1991). A
very bright and dusty red object is also seen lying along the critical
curve ($z=0.81$), as well as several lensed background elliptical
galaxies which are redder than the cluster E/SO's due simply to their
larger k-corrections (Frye \etal 1998).  The multiply lensed high
redshift images reported here are visible as relatively red and thin
arcs with similar substructure along their lengths, centered on a
bright cluster elliptical. Removal of this elliptical by ellipse
subtraction reveals 4 similar images forming a stretched set of four
arcs, as indicated in Figure 1b.

  Targets for multislit spectroscopy were then selected from the HST
field and supplemented by others registered on larger CFHT images, in
a general redshift survey of lensed red galaxies selected with
V-I$>$1.5 and I$<$24. Instrument flexure, fringing and the large
quantity of spectroscopy requires a purpose built code for efficient
and accurate reduction. Details of the data reduction and the redshift
survey are presented by Frye \etal (1998), complimenting the brighter
limited survey of the main arcs by Bezecourt \& Soucail (1997).  The
spectra of the larger arcs, labeled N and S in Figure 1, were taken
normal to the long axes of the two bright arcs using the W. M. Keck I
telescope in August 1996 in one hour of exposure, using LRIS with the
300 line grating and a slit width of 1.5'', resulting in a resolution
of 15.6 \AA \ and dispersion of 2.4 \AA \ per pixel. These spectra are
flux calibrated and shown in Figures 2a and 2b. The slit width is
narrower than the length of these arcs, 5'' and 3'', and oriented near
perpendicular the their long axes, hence the spectra are not expected
to correspond to same spatial section of the source, and their spatial
locations are indicated in Figure 1c. Subsequent spectroscopy in
August 1997, using the same instrumental configuration (using Keck II)
but with slits aligned along the arcs, totaling 2.8hrs of
exposure, confirms the high redshift of these arcs through the
presence of metal absorption lines. In addition, Ly$\alpha$
emission is seen at the Northern end of the Northern arc. A third arc
labeled E in Figure 1 was also observed to investigate the lens
geometry and found to share the same redshift as the two bright arcs,
its spectrum is shown in Figure 2c. The fourth image, W, is too faint
for useful spectroscopy.

\section{Spectral Analysis}

  Using the flux calibrated spectra of the N and S arcs we find that
their continua are very well fitted by a model B3 star (Kurucz 1993),
modified by the opacity expected from the Lyman-forest based on QSO
spectra, shortward of Ly$\alpha$ (as parameterized by Madau 1995),
as shown in Figure 2a,b. The lack of O-stars makes this object unusual
in comparison to the spectra of other high redshift galaxies. Taken at
face value this might simply imply that we are not catching this
galaxy in the throws of star-formation but shortly after, at
$10^6<t<3\times10^7$yrs, when the O-stars have disappeared.  But dust
reddening of an O-star population could also be invoked to explain the
continuum slope. The lack of flux shortward of restframe 1000\AA \ is
consistent with the expected fall off of the B-starlight following the
Wein tail.  However, since this fall off lies close to the Ly-limit at
912\AA, and better data is needed to discriminate between reddened
O-starlight as the source of the continuum emission and unreddened
B-stars. Of course the presence of Ly$\alpha$ emission requires a
source of FUV photons, but since a few O-stars generate a large
equivalent-width of emission, then a B-star dominated
spectrum is not problematic.

  A combined spectrum of the N and S arcs is also shown in figure 2c
(unfluxed), from data obtained with slits aligned along the major axes
of the arcs. This leads to an accurate redshift based on metal
absorption lines, SII 1260\AA, OI 1302\AA, SiII 1304\AA, CII 1336\AA,
SiIV 1394\AA,1403\AA, and SiIV 1403\AA at $z=4.04$. A foreground
absorption line is found covering both the N and S arcs, at
$\lambda=6454$\AA, most likely corresponding to MgII 2802\AA, at
$z=1.3$. The metal lines are clearly offset in wavelength from the
centroid of the Ly$\alpha$ emission which lies 300Km/s redward of the
metal lines (Figure 2). This redward shift is evident from inspection
of all high quality spectra of distant galaxies (Steidel \etal
1996a,b, Steidel \etal (1997),Trager \etal (1997), Franx, \etal (1997),
Lowenthal \etal (1997) and is seen in locally in the UV spectra of
starburst galaxies (Lequeux \etal 1996, Legrand \etal 1997), as
pointed out by Franx \etal (1997). Lequeux \etal (1996) explain this
shift as arising from back-scattered Ly$\alpha$ photons from the HI
shells around expanding HII regions, so that the Doppler redshift of
far-side emission is cleared of foreground absorption within the
galaxy.

 From the spatially resolved data it is clear that the Ly$\alpha$
emission is confined to the northern ends of both the N and S arcs.
The non-appearance of Ly$\alpha$ emission in the spectrum of the
Northern arc, Figure 2a, is explained by comparison with parallel slit
data, which showed our original slit location fell to the south of the
Ly$\alpha$ emission (Figure 1b). At this location damped HI absorption
is found with a column $\sim 3\times10^{21}n_{HI}$, sufficient to dilutely
spread any resonantly scattered Ly$\alpha$ emission over an
undetectably large wavelength range. The damped absorption is much
wider than for B-stars, hence such stars do not contribute to the
determination of the saturated HI column.  The presence of damped
absorption would seem to strengthen the relation advocated by Wolfe
(1988), between galaxies and high-redshift damped absorbers in
general.

\section{Lens Model}

  It is relatively straightforward to reproduce the basic properties
of the lensed image configuration. For generating four images
stretched along the major axis of the central elliptical member
requires a significant deflection from the cluster potential, so that
the elliptical member galaxy must lie close to the critical curve of
the cluster for a source at $z=4$. Figure 1c shows a reasonable model
of the data, where the deflection field is calculated numerically.
The critical curve of the cluster joins with that of the elliptical
galaxy at 30'' from the cluster center.  The absolute image separation
is set mainly by the cluster mass, and the relative separations of the
N-S over the E-W images pairs constrains the ratio of
elliptical/cluster masses to be $\sim 0.03$, depending somewhat on the
ellipticity of the cluster. We fix the ellipticity of the elliptical
galaxy to equal that of its light profile, $b/a=0.7$, resulting in a
relatively large ellipticity, $b/a=0.5$, for the cluster mass so
that the largest arc, N, can lie closer to the central elliptical than
arc S. In addition, an offset of $\sim 30$ degrees is needed between
the major axis of the cluster east of a line connecting the cluster
center through the elliptical member. This asymmetry also helps
somewhat to reproduce the straightness of the N and S arcs. For fixed
ellipticities, the relative image intensities are most sensitive to
the gradient of the cluster mass profile, preferring a shallower than
isothermal projected slope, $\theta^{-0.8}$, at the critical radius,
helping to generate a fairly lengthy arc E as observed.

  This result is shown in Figure 2 and displays the parity of the
images for a simple circular source, where the lighter and darker
regions of the arcs are adjusted to correspond to the regions of the
observed Ly$\alpha$ emission and the stellar continuum respectively,
matching the parallel slit spectroscopy. Also shown is the
magnification field, revealing the critical curve, demonstrating the
asymmetry required of the mass distribution. The source magnification
in this model is $\sim 20$. Taking the Northern arc alone and
subtracting its magnification from the observed magnitude
$I_{AB}=23.0$, yields $I_{AB}\sim25.5$ for the intrinsic apparent
magnitude, where the largest uncertainty is the slope of the mass
profile - the flatter the profile the larger the magnification. The
observed magnitudes of the the other arcs labeled S,E,W are
$I_{AB}$=23.6,24.5,26, respectively. The angular diameter of the
source in this model is $\sim 0.2 \arcsec$, or between 0.7kpc/h,
($\Omega$=1) and 1.4kpc/h ($\Omega$=0), assuming the source is roughly
symmetric. This width is consistent with the poorly width of the in
arcs in the HST images (Figure 1b).

  Our simple lens model is similar but not as extreme as those
explored in the analysis of the notorious straight arc in this cluster, which
has proven difficult to reproduce (Kassiola, \etal
1992). Interestingly this arc lies very close to
our high redshift system, shown in Figure 1, and is well resolved in
HST images (see Figure 1b where the adjacent elliptical is
removed). This difficulty is alleviated somewhat by the results of our
spectroscopy, which show unambiguously that component A (in the
notation of Pell\'o \etal 1991) lies at higher redshift, $z=1.033$,
than component C, which we confirm to be at $z=0.913$, and indicated
in Figure 1b.  Hence in our lens modeling we do not invoke an
additional dark mass (Kassiola \etal 1992) but instead prefer a large
ellipticity for the cluster mass distribution. Having said that,
it seems clear that this model does not produce the alignment of the
nearby N and S arcs, indicating the need for additional tangential
shear.

\section{Discussion and Conclusions}

  The inferred small size of our distant galaxy fits into the general
picture now emerging of galaxy formation by a piecemeal growth of
sub-galactic sized objects. The deepest images of the sky display a
surprisingly high surface density of tiny galaxies (Williams etal 1996)
and may be interpreted in a model-independent way as evidence for
growth on small scales at least (Bouwens, Broadhurst \& Silk, 1997,
BBS). Little is known about the process by which galaxies emerge from
the neutral epoch although clearly the bulk of this transition takes
place before $z\sim 1$, below which the density of L$>$L$^*$ galaxies
varies little (Lilly \etal 1995, Ellis \etal 1996). The Steidel population of
Ly-limit selected galaxies found in the range 2.5$<$z$<$3.5 (Steidel
{\etal} 1996a,1996b, Lowenthal, {\etal} 1997) have compact sizes and
display far more sub-structure than expected of redshifted spiral
galaxies (Bouwens \etal 1997). Individual star-formation rates are
uncertain, but several 10's of solar masses per year have been
inferred for the brightest cases (Pettini \etal 1997).  Their evolution and
relation to low-redshift galaxies is still unclear. By integrating
their UV luminosities, Madau \etal (1996) infer a steep decline in the
integrated rate of star-formation for z$>$2, relying on little
evolution in the shape of luminosity function. This steep decline is
unnatural in hierarchical models without strong `feedback'
(Cole \etal 1994). A simpler explanation for the behaviour of this
integral, in the context of a hierarchical scheme, results from the
continuous decrease in mean galaxy size and hence luminosity expected
with increasing redshift, naturally resulting in fewer galaxies
detected above the limiting luminosity. Indeed, flat or {\it
increasing} mean star-formation rates at z$>$2 may be accommodated by
the the observed dropout populations (Bouwens \& Broadhurst 1997)
in the range 2$<$z$<$6.

  Lensing by clusters, we have learned, provides a means to access the
highest redshift galaxies, aided by fortuitously large
magnifications. Fluxes enhanced by magnification effectively increase
the magnitude limit, allowing us to sample to higher $z$ behind the
cluster than in the field to a fixed flux limit.  However it should be
noted that the utility of lensing is compromised by the very
magnification, $\mu$, one would take advantage of.  The surface
density of sources is reduced by the expansion of sky area, at a rate
$1/\mu$, with each galaxy magnified by $\Delta m=2.5\log\mu$, so that
power-law counts transform via $N(<m) \propto \mu^{2.5\gamma-1}$, 
independent of the details of the mass distribution,
generating a net {\it reduction} for faint galaxies where $\gamma
<0.4$.  This implies a good estimate of the magnification is important
both for determining intrinsic luminosities and for estimating the
true surface density of source galaxies.

  The lensed galaxy presented here provides interesting information on
one of the earliest known optically selected galaxies. The cluster
magnification has fortuitously afforded us detailed spectral and
spatial information of a very faint and presumably otherwise typical
example of a galaxy at z=4.0. Subsequently we have obtained higher
resolution longslit spectra revealing spatial and spectral detail of
the Ly$\alpha$ and metal lines along the lengths of the N and S arcs
(Frye \etal 1998) and also near-IR Keck images useful for exploring
the role of dust and for a good estimate of the luminosity (Bunker\etal 1997).  The rarity of luminous high redshift galaxies and the
requirement of a high magnification for useful spectroscopic follow-up
means that such searches will be slow, but well rewarded by detailed
information on galaxies at otherwise inaccessibly early
times.

\subsection{Figures}

\begin{figure*}[t] 
\begin{center}
\leavevmode
\epsfxsize=6.5inch
\epsfysize=5inch
\epsffile{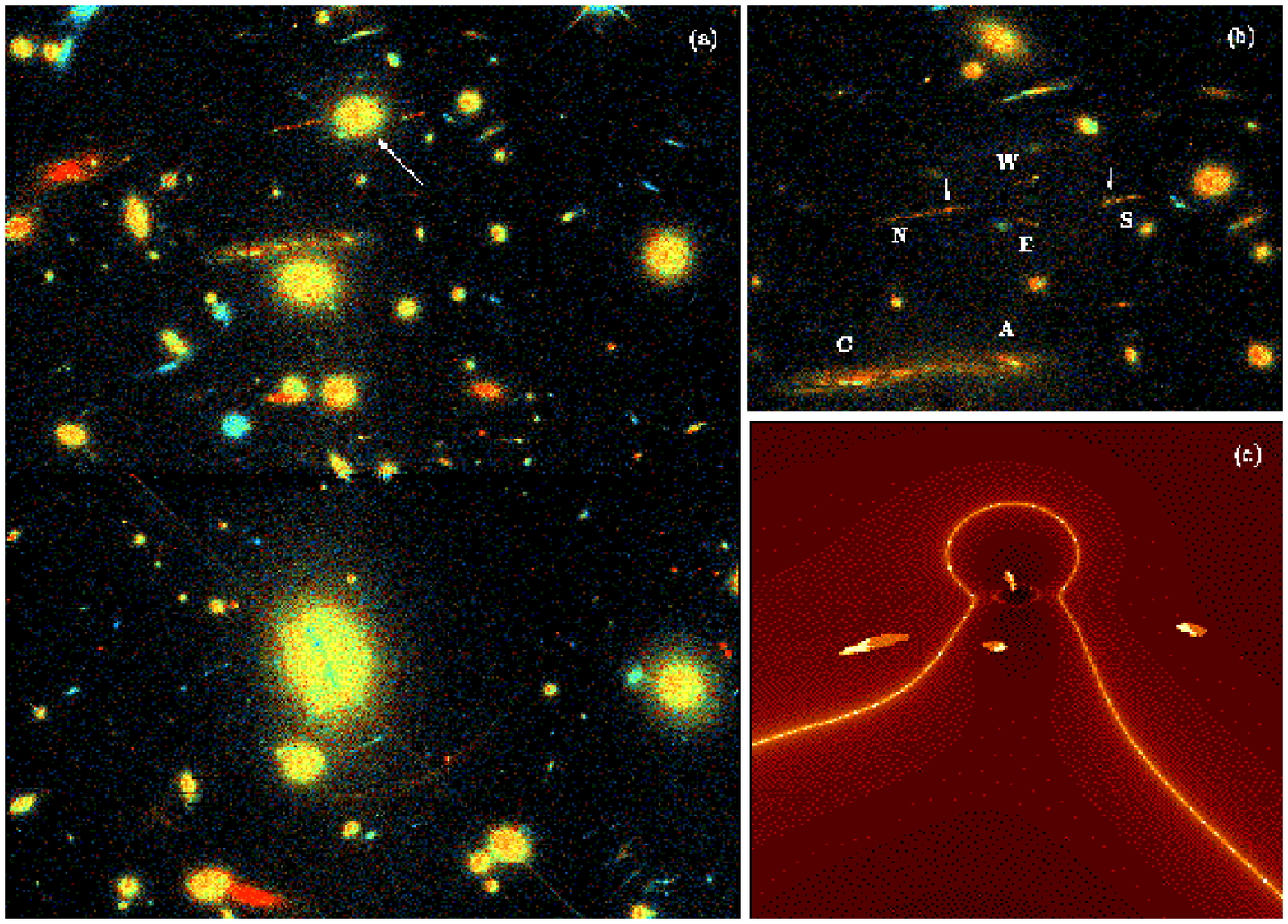}
\caption{Panel (a) shows the many red and blue arcs around the center
of A2390, with the elliptical galaxy centered on the z=4.04 arcs
indicated. (b) A blow up of the region containing the 4 lensed imaged of
the z=4.04 galaxy, with the central elliptical galaxy removed. The
locations of the slitlet positions corresponding to figures 2a,b are
indicated. Also shown is the ``straight arc'' which spectroscopy
resolves into two galaxies: component A, at z=1.033 and component C,
at the known redshift, z=0.913. (c) A simple lens model for the source
showing the importance of both the central elliptical and the cluster
potential in creating the image configuration. The parity of the
images is also also displayed. The brighter part of the images is 
chosen to match the spatial location of the observed Ly$\alpha$ 
emission in the N and S
arcs. The images are overlayed on the magnification distribution
showing clearly the bright asymmetric critical curve.}
\end{center}
\end{figure*}

\begin{figure*}[t] 
\begin{center}
\leavevmode
\epsfxsize=5.25inch
\epsfysize=6inch
\epsffile{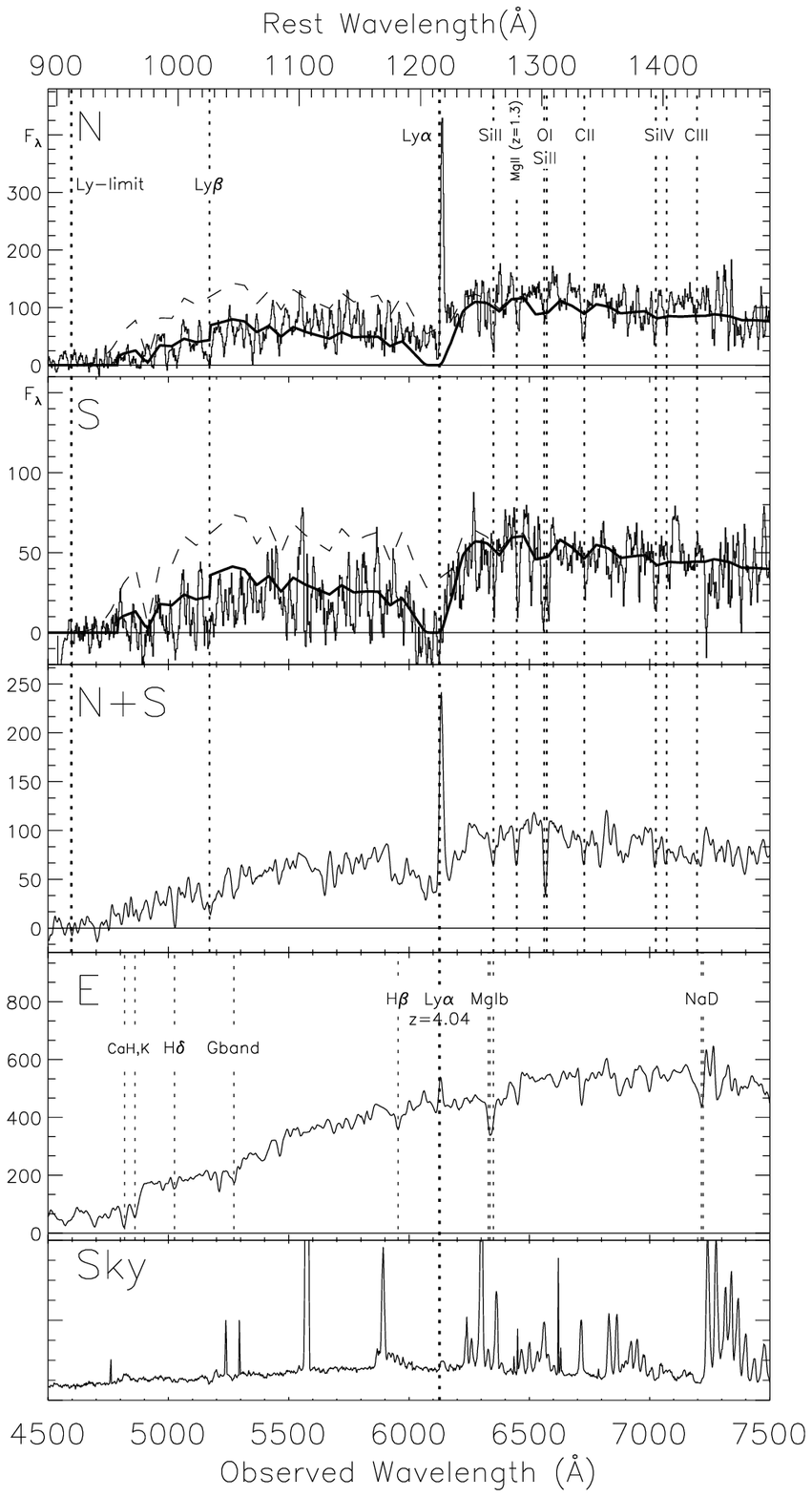}
\caption{{\bf (2)} Keck spectra of the arcs at z=4.04. The upper 
two panels show fluxed spectra of the Northern and Southern arcs
taken with slits aligned normal to the long axes of these arcs.
Overlayed is the best fitting B3-star continuum,
with the opacity of the forest included, together with 
damped Ly$\alpha$ absorption at the galaxy redshift, 
but no reddening. The middle panel 
shows the sum of both these arcs with slits aligned along their lengths,
revealing clearly metal lines at z=4.04, on which the redshift 
is based. These lines lie blueward of the centroid of the
Ly$\alpha$ emission by 300km/s, as discussed in the text. Below this is
shown a spectrum centered on the Eastern arc, revealing 
Ly$\alpha$ emission and the central
elliptical (z=0.223). The sky is shown for comparison at the bottom of the 
panel.
}
\end{center}
\end{figure*}

\acknowledgments

  We thank Hy Spinrad, Art Wolfe, Genevieve Soucail, Richard Ellis,
and Rosa Pell\'o and Fred Courbin for useful conversations. We also thank Art Wolfe
for his Ly$\alpha$ profile generating code and also Joe Morris and
Andrea Somer for help with the HST reductions.  TJB acknowledges an ST
grant NASA grant XYZ.

\fontsize{10}{14pt}\selectfont


\begin{references}

\reference {}Bouwens, Broadhurst \& Silk 1997 Astro-ph/9710291 

\reference {}Bouwens \& Broadhurst 1997 (in prep)

\reference{}Bezecourt, J. \& Soucail, G. 1997, A\&A, 317, 661

\reference{}Bunker, A., Moustakas, L. A., Frye, B. L, Broadhurst
T.J., Davis. M, \& Spinrad, H. 1998, {\sl The Young Universe, Rome}

\reference{}Cole, S., Aragon-Salamanca A, Frenk, C.S., Navaro, J.F., Zepf, S.E., 1994 MNRAS 271,781

\reference{}Dressler, A. \& Gunn, J. E. 1992, ApJS, 78, 1

\reference{}Ellis, R. S., Colless M., Broadhurst T.J., Heyl, J., Glazebrook, K., 
1996 MNRAS 280,235

\reference{}Ebbels, T. M. D., Le-Borgne, J.-F., Pell\'o, R., Ellis, R. S.,
Kneib, J.-P., Smail, I., \& Sanahuja, B. 1996, MNRAS, 281, L75

\reference{} Franx, M.,Illingworth,
G.D.,Kelson,D.,VanDokkum,P.G.,Pieter,G.,Tran,K., 1997 ApJ, 486,L75

\reference{}Frye, B., Spinrad, H., Broadhurst T.J., Bunker A., 1998 {\it in prep}

\reference{}Frye, B., Broadhurst T.J., Guhathakurta R, 1997 {\it in prep}

\reference{}Kassiola, A., Kovner, I. \& Blandford, R. D. 1992, ApJ, 396, 10

\reference{}Kneib, J.P.,Ellis, R.S.,Samil, I., Couch, W.J., Sharples, R.M., 1996, ApJ
. 471 643

\reference{}Kurucz, R. L. 1993, CD-ROM, Cambridge, MA: Smithsonian
Astrophysical Observatory, c1993

\reference{}Lowenthal, J. D., Koo. D. C., Guzman, R., Gallego, J., Phillips,
Faber, S.  M., Vogt, N. P., Illingworth, G. D., \& Gronwall, C. 1997, 
ApJ, 481, 673

\reference{}Lagrand, F.,Kunth,D.,Mas-Hesse, J.M.,Lequeux,J. 1997 A\&A, 326, 929

\reference{}Lequeux, F.,Kunth,D.,Mas-Hesse, J.M.,Sargent,W.L.W.. 1995 A\&A, 301, 18L

\reference{}Lilly, S.J., Tresse, L., Hammer, F., Crampton, D., Le Fevre, O., 1995 
ApJ 455,108

\reference{}Madau, P., Ferguson, H. C., Dickinson, M. E., Giavalisco, M.,
Steidel, C. C ., and Fruchter, A. 1996, MNRAS, 283, 1388

\reference{}Madau, P. 1995, ApJ, 441, 18

\reference{}Mellier, Y. 1989, in {\sl Clusters of Galaxies}, ed. M. Fitchett(Baltimore:  Space Telescope Science Institute)

\reference{}Pell\'o, R., Sanahuja, B., Le Borne, J.-F., Soucail, G., \&
                  Mellier, Y. 1991, ApJ, 366, 405

\reference{}Pettini, M., Steidel, C., Adelberger, K.,L., Kellogg, M., Dickinson, M., Giavalisco, M., 1997 {\sl Origins} eds Shull \etal (ASP Conference Series)

\reference{}Seitz, C, Kneib, J.-P., Schneider, P. \& Seitz, S. 1996,
A\&A,314,707

\reference{}Seitz, S, Saglia, R.-P., Bender, R. Hopp. U. Belloni, P., 
Ziegler, 1996, Astro-ph/9706023

\reference{}Steidel, C. C., Giavalisco, M., Pettini, M., Dickinson, M., \&
 Adelberger, K. L. 1996a, ApJ, 462, 17

\reference{}Steidel, C. C., Giavalisco, M., Dickinson, M., \& Adelberger,
 K. L. 1996b, AJ, 112, 352 

\reference{}Trager, S. C., Faber, S. M., Dressler, A. \& Oemler, A. 1997,
ApJ, 485, 92

\reference{}Yee, H. K. C., Ellingson, E., Bechtold, J., Carlberg, R. G. \&
Cuillandre, J.-C. 1996, AJ, 111, 1783

\reference{}Williams, R. E. \etal 1996 AJ 112,1335

\reference{}Wolfe, A.M., 1988 {\sl QSO Absortion Lines:Probing the Early
Universe, ed J.C. Blades, D.A. Turnshek, D.A.,\& C.A. Norman
(Cambridge:Cambridge Univ. Press}

\end{references}
\end{document}